\documentclass[aps,prb,reprint,twocolumn,showpacs,superscriptaddress]{revtex4-2}

\usepackage{amsmath}
\usepackage{graphicx}
\usepackage{lmodern}
\usepackage{amsmath}
\usepackage{color}
\usepackage{amssymb}
\usepackage{bm}
\usepackage{braket}
\usepackage{mathtools}
\usepackage{afterpage}
\usepackage{soul}

\renewcommand{\)}{\right)}
\renewcommand{\[}{\left[}
\renewcommand{\]}{\right]}

\DeclareMathOperator{\sgn}{sgn}

\newcommand{\be}{\begin{equation}}
\newcommand{\ee}{\end{equation}}

\usepackage[dvipsnames]{xcolor}
\usepackage[colorlinks=true]{hyperref}
\hypersetup{
    colorlinks=true,
    linkcolor=blue,
    citecolor=blue,
    urlcolor=blue
} 

\makeatletter
\def\maketitle{
\@author@finish
\title@column\titleblock@produce
\suppressfloats[t]
}
\makeatother

\makeatletter 
    
\renewcommand\onecolumngrid{% <<<<<<
\do@columngrid{one}{\@ne}%
\def\set@footnotewidth{\onecolumngrid}% <<<<<<<<<<<<<<<<
\def\footnoterule{\kern-6pt\hrule width 1.5in\kern6pt}%
}

\renewcommand\twocolumngrid{% <<<<<<
        \def\footnoterule{% restore rule
        \dimen@\skip\footins\divide\dimen@\thr@@
        \kern-\dimen@\hrule width.5in\kern\dimen@}
        \do@columngrid{mlt}{\tw@}
}%

\makeatother    
\begin{document}
\title{Gapless Fermionic Systems as Phase-space Topological Insulators: Non-perturbative Results from Anomalies}
\author{Taylor L. Hughes}
\affiliation{Department of Physics and Institute of Condensed Matter Theory, University of Illinois at Urbana-Champaign, Urbana, Illinois 61801-3080, USA}
\author{Yuxuan Wang}
\affiliation{Department of Physics, University of Florida, Gainesville, Florida 32611, USA}

\date{\today}

\begin{abstract}
We present a theory unifying the topological responses and anomalies of various gapless fermion systems exhibiting Fermi surfaces, including those with Berry phases, and nodal structures, which applies beyond non-interacting limit. As our key finding, we obtain a general approach to directly relate gapless fermions and topological insulators in phase space, including first- and higher-order insulators. Using this relation we show that the low-energy properties and response theories for gapless fermionic systems can be directly obtained without resorting to microscopic details. Our results provide a unified framework for describing such systems using well-developed theories from the study of topological phases of matter. 
\end{abstract}
\maketitle

\noindent\textbf{Introduction.---}
One of the most remarkable properties of topological insulators (TI) is their description in terms of quantized, detail-independent quantum anomalies~\cite{qi2008,witten2016}. In contrast, spectral, thermodynamical, and transport properties of gapless systems such as Fermi liquids~\cite{shankarRG,son2022}  typically depend on the details of the dispersion and interactions. However, there are properties of interacting, gapless fermionic systems (GFSs) that are independent of microscopic details. Perhaps the most well-known example is Luttinger's theorem for a Fermi liquid~\cite{luttinger,luttinger-ward, oshikawa2000,else2021}, which is a non-perturbative result that relates the total electric charge with the volume enclosed by the Fermi surface. 

In recent years, it has been increasingly appreciated that Luttinger's theorem itself has a topological origin~\cite{oshikawa2000}--- most directly, it naturally follows from a topological Chern-Simons term in \emph{phase space} (combining real and momentum space)~\cite{else2021,yizhuang}.  Indeed, phase-space topology has been utilized to classify gapped~\cite{qi2008} and gapless~\cite{bulmash,yizhuang} fermionic systems having various internal symmetries. Additionally, the topological nature of Fermi surfaces even places important non-perturbative constraints on transport.~\cite{hart-else-senthil} The power of phase space topology relies on an analogy between TIs, which harbor localized gapless excitations in position space, and GFSs which harbor gapless modes localized on submanifolds of momentum space, e.g., Fermi surfaces or nodal lines/points. 

While seminal work establishing this analogy was carried out in Refs.~\cite{bulmash,else2021,yizhuang}, in this work we obtain a general recipe for mapping GFSs to TIs in phase space. As a direct consequence of the phase-space approach, we obtain the low-energy properties, of both the bulk and the boundary, without relying on microscopic details. Importantly, such an approach yields a unified framework to understand a wide variety of GFSs including Fermi surfaces enclosing Berry phases and  topological semimetals. From the quantum anomalies of the phase space TI, we unify various response theories for these GFSs and generate results that act as new types of Luttinger theorems that remain valid in the presence of interactions. Our work also sheds light on the classification of a variety of GFSs.~\cite{horavas,yizhuang}

\noindent\textbf{General Approach.---}
\label{sec:summary} 
Our approach will be to associate a phase space TI (PSTI) having total phase-space dimension $D$ (not counting time) to a given GFS.  Then, we will use the known results of the topological response theories of TIs to describe properties of the GFS. \footnote{{We emphasize that this correspondence is limited to their \emph{topological} properties. For example, interactions in GFSs are nonlocal in $k$-space, leading to important differences with boundary states of (real-space) TIs. However, the topological properties are rooted in emergent quantum anomalies of the low-energy theory, which are robust against interactions~\cite{wangsau2024}.}} Let us denote the spatial dimension of the given GFS as $d$, and the co-dimension of the gapless-fermion submanifold in momentum space as $c$. While gapped in the ``bulk," a PSTI has gapless modes on boundaries (or defects) having co-dimension $C.$ While nontrivial surface states of first-order TIs correspond to $C=1$, it is also known that TIs can host nontrivial gapless states on boundaries/defects having $C=2$ (or larger values), such as in vortex cores, flux defects,~\cite{teo-kane} or the corners or hinges of a second-order TI.~\cite{Hughes_2017,Benalcazar_2017,Schindlereaat_2018,Shapourian_2018,Piet_2017,Piet_2019,Eslam_2018,Das_Sarma_2019b,Katsunori_2019,Zhongbo_2019b,Dai_Xi_2019,Vladimir_2019,Qiang_Wang_2018,Fulga_2018,Sun_2020,Penghao_2020,Piet_2018,Ezawa_2018,Fan_2019} The required dimension $D$ of the PSTI can be determined via the following counting. We need $d$ spatial dimensions, and also $d-c$ momentum coordinates to parameterize the gapless submanifold. In addition, we need $C$ momentum coordinates to represent the ``bulk" directions of the PSTI in momentum space. While the choice of the PSTI corresponding to a given GFS is not unique (i.e., the choice of $C$, and hence $D$ is not unique), we find that the correspondence between a GFS and a PSTI can be described by the following general relation
\be
D-C=2d - c.
\label{eq:1}
\ee
 We note that while the choice of $D, C$ are not unique, there is a minimal case where $C$ is chosen to be as small as possible. We list a few examples in Table \ref{tab:1}, and discuss several of them in detail below.  As we will show, the topological response theory for PSTIs in the presence of the appropriate background gauge field configuration gives the correct description of the robust features of the GFS. 

\begin{table*}[t]
  \caption{Examples for the correspondence between gapless fermionic systems and phase space TIs.}
  \label{tab:1}
  \begin{ruledtabular}
  \begin{tabular}{cccccc}
Gapless fermionic system & $d$ & $c$ & Phase space TI & $D$ & $C$ \\
    \hline
        1d Fermi surface    & 1   &  1  &2d QHE&2&1  \\ 
    2d Fermi surface/doped Dirac semimetal~\cite{SM}    & 2   &  1  &4d QHE&4&1  \\ 
        2d Dirac semimetal     & 2   &  2  &4d QHE/3d TI&4/3&2/1  \\ 
    3d Fermi surface/doped WSM    & 3   &  1  &6d QHE&6&1  \\ 
         WSM$_2$    & 3   &  3  &6d QHE/4d QHE&6/4&3/1  \\
    % Minimal WSM    & 3   &  3  &4d QHE&4&1  \\
    $S_4$-symmtric WSM   & 3  & 3   & 5d HOTI  & 5& 2 \\
          \end{tabular}
  \end{ruledtabular}
\end{table*}
 
Before we move on to some examples let us note some immediate consequences of Eq.~\eqref{eq:1} for the classification of the stability of GFSs. It is well known from Ref.~\cite{teo-kane} that the topology of a \emph{gapped} Hamiltonian is classified according to the difference between the position-space dimension $d$ and its (effective) momentum-space dimension. For a boundary having co-dimension $C$ in $(D-d)$-dimensional momentum-space, the effective momentum-space dimension for classification purposes is $(D-d)-(C-1)$~\cite{teo-kane}. 
Hence, the classification of the PSTI and, via bulk-boundary correspondence, the GFS, depends only on, using Eq.~\eqref{eq:1}, $d - [(D-d) - (C-1)] =c-1$, i.e., the classification depends only on the co-dimension of the gapless region of the GFS in momentum space.~\cite{chongwang2}
This fact has been recently appreciated in the classification of interacting Fermi surfaces ($c=1$) having various internal symmetries~\cite{yizhuang,horavas}, and we can use our relation to further extend the classification for all GFS's using known results for TI classification  (fermionic SPTs)~\cite{qi2008,ryu2010,kitaev2009,teo-kane,kapustin2014,kapustin2015}. While we postpone a  detailed study of the interacting stability classification to future work, we will discuss an example below for Weyl semimetals.

\noindent\textbf{Fermi surfaces.---}
We begin by considering Fermi surfaces ($c=1$) of systems having spatial dimension $d=1$. As shown in Fig. \ref{fig:1}(a), the minimal correspondence for a GFS in 1d is a Chern insulator in $D=2$ phase space where the 1d chiral Fermions at the Fermi surface (FS)  can be regarded as edge states of the Chern insulator, i.e., edge states that are localized at a fixed momentum instead of a spatial boundary. The bulk phase-space theory that cancels the chiral anomaly of the FS~\cite{chongwang1, chongwang2} is given by
\be
S=\frac{1}{4\pi}\int_{w,x,t} \mathcal{A}\wedge d\mathcal{A},
\ee
%While in Ref.~\cite{...} was obtained from the chiral anomaly of the low-energy theory, our derivation can be viewed as from the `bulk'  in phase space.
where $w\sim k_x$, and  $\mathcal{A}$ is the background gauge field in phase space. In particular, we expand the 1-form gauge field $\mathcal{A}$ as follows:
\be
\mathcal{A} = eA + \mathfrak{a}+ w \mathfrak{e}^x + w^2  \mathfrak{e}^{xx} + w^3  \mathfrak{e}^{xxx} + \cdots,
\ee
where $A$ is the electromagnetic gauge field, $\mathfrak{e}^{abc\ldots}$ are the translation gauge field and generalizations thereof, all of which have background values that live only in position space, and $\mathfrak{a}$ is the Berry connection, which, for our purposes, lives only in momentum space.  Importantly, one needs to properly encode the information that some phase space coordinates, e.g., $x$ and $w\sim k_x,$ should not commute. This noncommutativity can be restored by adding a background gauge flux $F_{x,w}=1$ via a gauge choice $\mathcal{A}_x= w,$ and projecting to an effective Landau level, analogous to noncommutative geometry emerging in quantum Hall physics.~\cite{bellissard1994} Remarkably this immediately implies that the translation gauge field has a background value 
$\mathfrak{e}^i_\mu=\delta^i_\mu$ stemming from the fundamental relation that momentum translates position. The background values for generalized translation gauge fields $\frak{e}^{ab...}$ are zero, but they may be turned on as probe fields.

\begin{figure}[t!]
    \includegraphics[width=1.\columnwidth]{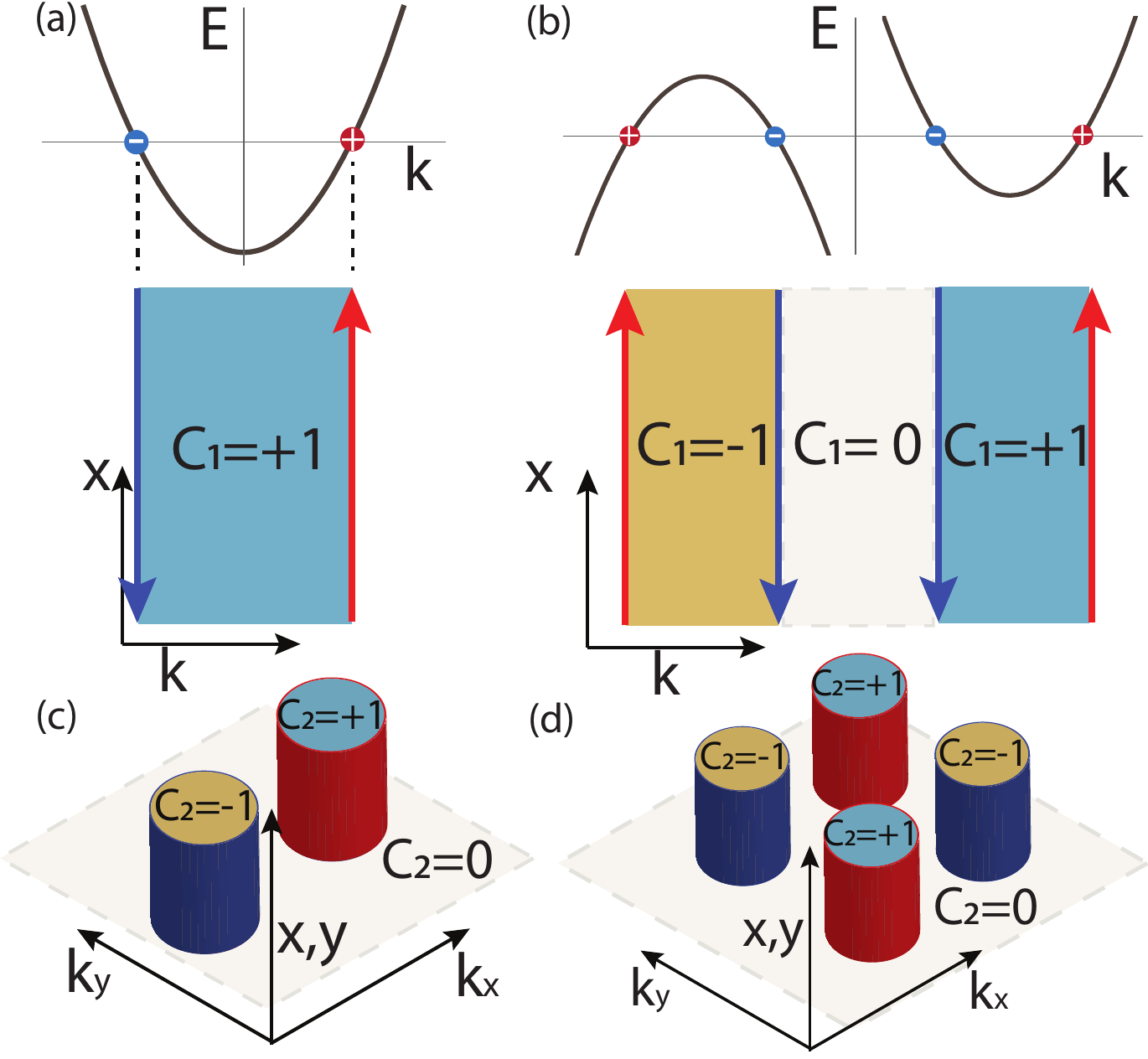}
    \caption{Illustration of PSTIs corresponding to  1d FSs in (a),(b) and 2d FSs in (c), (d). Chern numbers $C_{1,2}$ with positive/negative signs correspond to electron/hole-like pockets. In (a), (b), $D=2,C=1,d=1,c=1$, and in 
    (c), (d), $D=4,C=1,d=2,c=1$. As discussed in the text, in (b,c,d) the system has a vanishing (i.e., compensated) density and trivial Luttinger theorem relation, but satisfies additional generalized Luttinger theorems for higher powers of momentum.}
    \label{fig:1}
\end{figure}

Using this response action in the presence of the background configuration for $\mathfrak{e}^a$ immediately yields several non-perturbative results for various powers of the total momentum $P_x$ (the zeroth power being the total particle number $N$):
\begin{align}
N =& \int_{w,x} \frac{\delta S}{e\delta A_0} = \frac{1}{2\pi}\int dw dx\, \partial_w  \mathcal{A}_x = \mathcal{P} \frac{L}{2\pi},\nonumber \\
\langle P_x \rangle=& \int_{w,x} \frac{\delta S}{\delta \mathfrak{e}^x_0} = \frac{1}{2\pi}\int dw dx\,  w \partial_w  \mathcal{A}_x = \frac{\mathcal{Q}}{2}\frac{L}{2\pi}, \nonumber \\
\langle P_x^2 \rangle=& \int_{w,x} \frac{\delta S}{\delta \mathfrak{e}^{xx}_0} = \frac{1}{2\pi}\int dw dx\,  w^2 \partial_w  \mathcal{A}_x = \frac{\mathcal{O}}{3}\frac{L}{2\pi},
\label{eq:2}
\end{align}
where we have used $\mathfrak{e}^x_x=1$ as a background value, and all other components of $\mathcal{A}$ are chosen to vanish.
Here $\mathcal{P} = \int dw = \sum_{a} \chi_a k_F^{(a)}$ is the momentum-space dipole moment formed by all of the Fermi points $a=1,2,... N_F$, where the $\chi_a$ are their chirality, and the $k_F^{(a)}$ are their corresponding Fermi momenta. The quantities  $\mathcal{Q}$ and $\mathcal{O}$ are quadrupole and octupole moments of the Fermi points, and they are gauge-invariant only if all lower moments vanish. For example, this implies that $\mathcal{Q},$ and hence $\langle P_x\rangle,$ is well-defined and independent of the origin of momentum space only if $\mathcal{P}=0$ and $\chi=\sum_a \chi_a=0$ as is the case for the configuration in Fig. \ref{fig:1}(b).~\cite{hirsbrunnerDubBurHug2023}. 

The first equation in \eqref{eq:2} is simply the usual Luttinger theorem in 1d, while the subsequent equations are hierarchical generalizations of Luttinger's theorem that apply when each of the lower quantities vanish. For example, for a compensated semimetal that has a vanishing total density (which we can associate to a PSTI heterostructure as in Fig. \ref{fig:1}(b)), our result predicts a relation between the total momentum and the quadrupole moment of the Fermi points. Furthermore, if the total momentum also vanishes, the total squared momentum is related to the octupole moment of the Fermi points, and so on. These results can be regarded as generalized Luttinger relations in the sense that they relate short-distance quantities such as charge and momentum with the low-energy properties of the Fermi surfaces.  %Additionally, we can see from the Green function relations {\bf{TLH: Put in $N=G^{-1}dG$ and $P=k G^{-1}dG$ formulas. Maybe also mention that zeros can mess up Luttinger count, but even if they don't they can mess up the momentum count etc.}} that the results are well-defined if the lower moments vanish (i.e., independent of $k\to k+\delta k$) and perturbatively stable in the presence of interactions. 

The generalized Luttinger theorems can be straightforwardly extended to higher dimensions. As an example, we can consider compensated semimetals in 2d having vanishing total density and broken or preserved inversion symmetry as shown in Fig.~\ref{fig:1}(c),(d) respectively. To be explicit, let us assume circular Fermi pockets with radius $k_F$ centered at $(k_x, k_y)=(\pm K_x,0)$ for Fig. \ref{fig:1}(c), and around $(k_x, k_y)=(\pm K_x, \pm K_y)$ for Fig. \ref{fig:1}(d). Both of these gapless systems can be understood as PSTIs having $D=4$ according to Eq.~\eqref{eq:1} where $d=2, c=1, C=1.$ The $D=4$ phase space is parameterized by $(w\sim k_x, v\sim k_y, x, y)$, where inside each electron or hole pocket we have a TI having a second Chern number $C_2=\pm 1$ respectively. Indeed, in Ref.~\cite{yizhuang} the authors showed that  the surface states of these 4D Chern insulators are the low-energy states near the 2d Fermi surfaces. The PSTI response theory is given by
\be
S=\sum_{a=1}^{N_{\rm FS}}\frac{C_{2a}}{24\pi^2}\int_{(w,v)_a,x,y,t} \mathcal{A}\wedge d\mathcal{A}\wedge d\mathcal{A},
\label{eq:5}
\ee
where $N_{\rm FS}$ is the number of Fermi surface components and $C_{2a} = \pm 1$ is the second Chern number for each electron or hole pocket respectively. This action has been proposed to describe the LU(1) 't Hooft anomaly for a 2d Fermi liquid in Refs.~\cite{else2021,ma2021emergent,else2023holographic}, and analogous to Eq.~\eqref{eq:2}, Eq.~\eqref{eq:5} straightforwardly leads to the 2d Luttinger theorem~\cite{yizhuang}. 

We can now show some additional consequences of this response by considering the FS's in Fig. \ref{fig:1}(c) (Fig. \ref{fig:1}(d)). The FS components are arranged in a dipolar (quadrupolar) pattern so we expect that the system will have a nonzero value of $\langle P_x\rangle$ ($\langle P_x P_y \rangle $). To extract this from the response we expand $\mathcal{A} = eA + w \mathfrak{e}^x +v \mathfrak{e}^y + w^2  \mathfrak{e}^{xx} + v^2  \mathfrak{e}^{yy} + vw  \mathfrak{e}^{xy} + \cdots$ and perform the integrals in the $w, v$ directions. We hence find the two expected responses for the dipolar and quadrupolar FS arrangements are captured by the following contributions to Eq.~\eqref{eq:5}
 \begin{eqnarray}
S_{\mathcal{P}}&=&\frac {\mathcal{P}_{i} V}{4\pi^2}\int_{x,y,t}  \mathfrak{ e}^{i}_0 \mathfrak{e}^x_x  \mathfrak{e}^y_y,\\
S_{\mathcal{Q}}&=&\frac {\mathcal{Q}_{ij} V}{4\pi^2}\int_{x,y,t} \mathfrak{ e}^{ij}\wedge \mathfrak{e}^x \wedge \mathfrak{e}^y,
 \end{eqnarray}
where  $V$ is the volume of the system in position space, $\mathcal{P}_{i} = \sum_a\int C_{2a} k_i^{(a)} dk_x^{(a)} dk_y^{(a)}$ is the dipole moment of the Fermi pockets, and $\mathcal{Q}_{ij} = \sum_a\int C_{2a} k_i^{(a)}k_j^{(a)} dk_x^{(a)} dk_y^{(a)}$ is the quadrupole moment of the Fermi pockets. For the configuration in Fig. \ref{fig:1}(c) $\mathcal{P}_x=2K_x(\pi k_{F}^2)$, and for the configuration in Fig. \ref{fig:1}(d) $\mathcal{Q}_{xy}=2K_x K_y(\pi k_{F}^2),$ both of which are simply the 1d formulae for the Fermi-point dipole and quadrupole of the Fermi-pocket centers, but multiplied by the area of the Fermi pockets. The dipole $\mathcal{P}_x$ is unambiguously defined because the total density vanishes in Fig. \ref{fig:1}(c), while $\mathcal{Q}_{xy}$ is well-defined because the density and $\mathcal{P}_i$ both vanish in Fig. \ref{fig:1}(d). We thus have
 \begin{align}
 \langle P_x\rangle=& \int_{x,y} \frac{\delta S_{\mathcal{P}}}{\delta \mathfrak{ e}^x_0}=\frac{V}{4\pi^2} \mathcal{P}_x,\\
 \langle P_x P_y\rangle =& \int_{x,y} \frac{\delta S_{\mathcal{Q}}}{\delta \mathfrak{ e}^{xy}_0}=\frac{V}{4\pi^2} \mathcal{Q}_{xy},
 \end{align} which are the 2d analogs of Eq.~\eqref{eq:2}.
 We note that if the system has $C_4$ or $C_6$ symmetry, $\mathcal{Q}$ has to vanish, and in that situation, the generalized Luttinger's theorem provides a non-perturbative relationship between expectation values of cubic powers of momentum and the octupole moment of the Fermi surface. We will discuss more details about these generalized Luttinger theorems in a future work~\cite{wanghughesnextpaper}.
 
\noindent\textbf{FS's with Berry Phases.---}We will now demonstrate that the phase-space topological response theory can straightforwardly incorporate Berry phase effects in GFSs. Two important examples are the Fermi surfaces of a doped 2d Dirac semimetal, which enclose a $\pi$ Berry phase, and the Fermi surfaces of a doped 3d Weyl semimetal (WSM) which enclose a Berry curvature monopole. We discuss the former in the Supplement~\cite{SM}, while here we consider a doped WSM having two spherical, hole-like FS's with radii $k_{F1}=k_{F2}\equiv k_F$ where each FS encloses a single Weyl point centered at $\vec k^{(a)}$ with chirality $\chi_a=\mp 1$,  $a=1,2.$  According to Eq.~\eqref{eq:1}, the corresponding PSTI to each WSM FS is a $D=6$ Chern insulator parameterized by $(w\sim k_x, v\sim k_y, u\sim k_z, x, y,z)$, having a third Chern number $\mathcal{C}_3=1$ enclosing a vacuum region in the $(w,v,u)$ subspace. Indeed, for this construction the gapless Weyl fermions are essentially a Witten-effect-like response to momentum-space magnetic monopoles in the PSTI (where $D=6$ and $C=3$).~\cite{witten1979,qi2008,witteneffectsolution} 

The $D=6$ PSTI response action is given by~\cite{SM}
 \be
S=\frac{1}{192\pi^3}\int_{w,v,u,x,y,z,t} \mathcal{A}\wedge d\mathcal{A}\wedge d\mathcal{A}\wedge d\mathcal{A}.
\label{eq:9}
 \ee Keeping up to only linear couplings to momentum, we decompose the gauge field as $\mathcal{A} = eA +  w\mathfrak{e}^x+v\mathfrak{e}^y +u\mathfrak{e}^z+ \mathfrak{a}_m$,
where  $\mathfrak{a}_m$ is Berry connection created by the two monopoles in $\vec k$-space; over a  surface enclosing one of the monopoles $\oint \nabla\times \mathfrak{a}_m = \pm 2\pi$. The background configurations of the Berry monopole and the translation gauge fields generate the response theory up to quadratic order in $A$:
\begin{align}
\!\!\!S=&\int_{\mathbb{R}^4} \left[ \frac{e\mathcal{ V}}{8\pi^3} A\wedge \mathfrak{e}^x \wedge \mathfrak{e}^y \wedge \mathfrak{e}^z -\frac{e^2\mathcal {P}_i}{8\pi^2} \mathfrak{e}^i\wedge A\wedge dA \right] ,
\label{eq:10}
\end{align}
where $(t,x,y,z)\in \mathbb{R}^4.$ The first term represents Luttinger's theorem in 3d, where $\mathcal{V}$ is the volume enclosed by the two FS's.~\cite{SM}
The second term is a response induced by $\mathfrak{a}_m$ describing the anomalous Hall effect $\sigma_{ij} = -\epsilon_{ijk} \tfrac{e^2}{4\pi^2}\mathcal{P}_k$.~\cite{chongwang1,hirsbrunnerDubBurHug2023} From Eq.~\eqref{eq:10} there also exists a momentum current response to the translation gauge field $\mathfrak{e}^i$ via $\mathfrak{j}_i^\mu\equiv \delta S/\delta \mathfrak{e}^i_\mu$, which carries  the $i$-th component of momentum  in the $\mu$-th direction. The total momentum is $\langle P_i\rangle=\int_{\mathbb{R}^3}\star\mathfrak{j}_i,$ and from Eq.~\eqref{eq:10} we obtain~\cite{wangsau2024}
\begin{align}
\!\!\!\!\!\partial_t \langle P_i\rangle dt %=& \int_{\mathbb{R}^3}  d\star \mathfrak{j}_i= \int_{\mathbb{R}^3}  d\star \frac{\delta S}{\delta \mathfrak{e}^i} \nonumber\\
=& \int_{\mathbb{R}^3} \[\frac{e\mathcal V}{16\pi^3}\epsilon_{ijk} F\wedge \mathfrak{e}^j \wedge \mathfrak{e}^k  - \frac{e^2 \mathcal{P}_i}{8\pi^2} F\wedge F \],
\label{eq:140}
\end{align}
where $F=dA$ is the background electromagnetic field. 
Using Luttinger's theorem, the coefficient of the first term in the second line of Eq.~\eqref{eq:140} is equal to the electric force $eNE_i$. The second term $\propto \vec E \cdot \vec B$  describes additional momentum generation via the chiral Landau levels (cLLs) of the Weyl fermions even at $\mathcal V =0$.~\cite{son-spivak}

The same result \eqref{eq:140} can be equivalently seen from an emergent anomaly of the PSTI boundary. To illustrate this connection, note that from the PSTI analogy, the  gapless fermions on the $a$-th FS exhibit an emergent charge anomaly, given by
$d\star j^{(a)}(\vec k_F^{(a)}) = \frac{e}{16\pi^3} F\wedge d(k_j \mathfrak{e}^j)\wedge d(k_l \mathfrak{e}^l)+\frac{e^2}{16\pi^3} F\wedge F\wedge \mathfrak{f}_m(\vec k_F^{(a)})$,~\footnote{By integrating over the FS, one obtains the chiral anomaly from the entire FS obtained in Ref.~\cite{son2012}.}
where $\mathfrak{f}_m = d\mathfrak{a}_m$ is the Berry curvature on the Fermi surface, and $\vec k_F^{(a)}$ is a Fermi momentum.
 The momentum current depends on the particle current and is given by $\mathfrak{j}^{(a)}_i \equiv \oint_{{\rm FS}_a}  k_{F,i}^{(a)}\, j^{(a)}(\vec k_F^{(a)})$,  and $\partial_t\langle P_i\rangle dt =\sum_a \int_{\mathbb R^3} d\star \mathfrak{j}_i^{(a)}$ evaluates to
\begin{align}
\!\!\!\!\!\partial_t\langle P_i\rangle dt=&\int_{\mathbb{R}^3} \frac{e\mathcal V}{16\pi^3}\epsilon_{ijl} F\wedge \mathfrak{e}^j \wedge \mathfrak{e}^l \nonumber\\
&+  \sum_a\oint_{{\rm FS}_a}  
k_{F,i}^{(a)}\,\frac{\mathfrak{f}_m(\vec k_F^{(a)})}{2\pi} \int_{\mathbb{R}^3}\frac{e^2}{8\pi^2} F\wedge F.
\label{eq:160}
\end{align}
{Matching  \eqref{eq:140} and \eqref{eq:160}} immediately leads to 
\be
\mathcal{P}_i = -
\sum_a\oint_{{\rm FS}_a}  
k_{F,i}^{(a)}\frac{\mathfrak{f}_m(\vec k_F^{(a)})}{2\pi},
\label{eq:170}
\ee
which agrees with a direct bulk evaluation.~\cite{SM} 
For linear dispersion around the FS this simplifies to $\mathcal{P}_i = \sum_{a=1}^2 \chi_a k_i^{(a)} $, i.e., the anomalous Hall effect $\sigma_{ij}$ is proportional to  the dipole moment of the Weyl points, independent of doping. 
%This result was previously known for the undoped case~\cite{chongwang1,hirsbrunnerDubBurHug2023}. 
The weak dependence on doping was first derived using a Green's function method~\cite{burkov2014}. Moreover, from the general formula Eq.~\eqref{eq:170}, $\mathcal{P}_i$ receives contributions from all FS points weighted by the  Berry curvature. That the anomalous Hall effect in 3d is a FS phenomenon~\cite{chen2017,huang2023} was first pointed out in Ref.~\cite{haldane2004}, but compared with earlier works, our result is established non-perturbatively~\footnote{{For interacting systems, the Berry curvature $\mathfrak{f}_m$ on the FS should be understood as defined via Green's functions~\cite{setty2023, subir2023}.}} from anomaly inflow.

%The last two terms of ~\eqref{eq:10} combine to generate a response localized on the FSs, and they vanish if there is no boundary in momentum space, i.e., if the system is gapped without a FS. We can see this since $A$ has components only in spacetime, so one of the derivatives in the 5-form $A\wedge dA\wedge dA$ must be with respect to a momentum. \YW{(YW: I am not sure this previous sentence is correct. I agree that one gradient is with respect to momentum, but it does not make the 5-form a total derivative in that momentum direction.)}
%{\bf{TLH: Need to clarify this notation. where $k_r^{(a)}\in K_F=(0,k_F)$ is range of the radial component of the momentum deviation from the respective Weyl point (assuming spherical FS's),  the origin of which connected in a ``bulk" with $W\equiv[k_x^{(1)},k_x^{(2)}]$}} 
%Indeed, these terms encode the chiral anomaly of the gapless fermions on each FS via the non-conservation of the electromagnetic current:
%\be
%\partial_\mu j^\mu = \int_{k_r^{(a)}\in K_F, w\in W} d\star j = -\frac{\star (F\wedge F)}{8\pi^2}\sum_a\chi_a,
%\ee
%\YW{(YW: This equation above doesn't make sense...)}
%where $F=dA,$ and the total chirality $\sum_a \chi_a=0$ for lattice models satisfying the Nielsen-Ninomiya theorem~\cite{nielsenninomiya}. %The integral in $w\in W$ represents a ``bulk" action that cancels the chiral anomaly from each of the two FSs.

 \begin{figure}[t!]
    \includegraphics[width=1.\columnwidth]{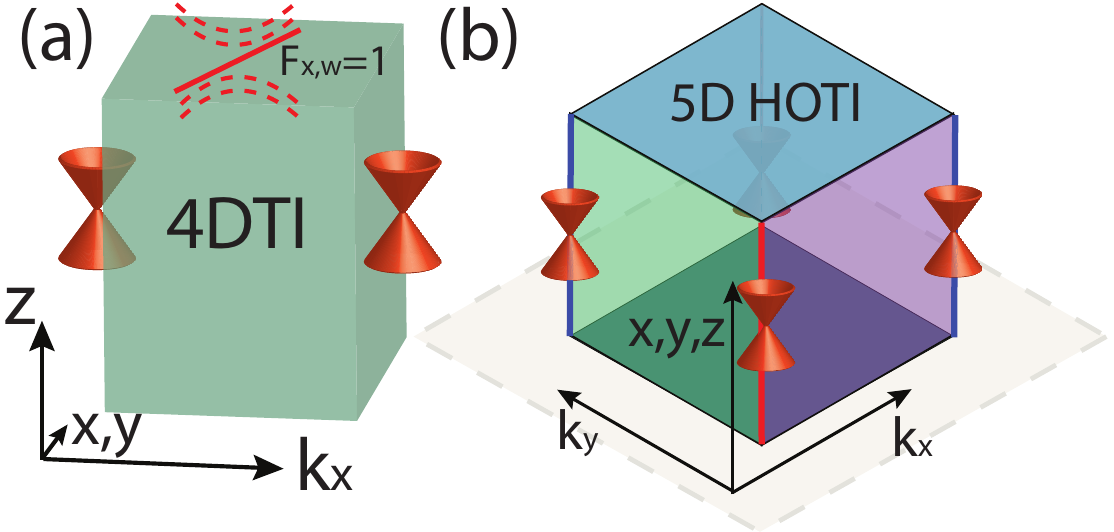}
    \caption{Illustration of PSTIs for WSM$_2$ and $S_4$-symmetric WSM. In (a), $D=4,C=1,d=3,c=3$, and we see bulk Weyl nodes on the surfaces in the $k_x$ direction while the $z$-surface modes are subjected to a phase space magnetic field $F_{x,w}$ that generates the Fermi arcs on the WSM surface. In 
    (b), $D=5,C=2,d=3,c=3,$ and bulk Weyl nodes appear on the hinges as domain wall states generated by opposing surface mass terms (purple and green masses have opposite signs). Weyl points on neighboring hinges have opposite chiralities.} 
    \label{fig:2}
\end{figure} 
  
\noindent\textbf{Weyl semimetal.---}
As one  shrinks the Fermi pockets to Weyl points,  the first term in Eq.~\eqref{eq:10} vanishes, and we are left with the expected anomalous Hall response from nodal Weyl points.  In this nodal, semi-metallic limit, instead of including all three momentum coordinates, a simpler, alternative description of the WSM having two Weyl points (WSM$_2$) is constructed by viewing the Weyl points as surface states of a 4d Chern insulator as shown in Fig. \ref{fig:2}(a). In this description, the phase space TI has dimension $D=4$ and $C=1$. For $\vec k^{(1,2)}=(\pm K_x,0,0)$, the ``bulk" PSTI response action is described by~\cite{chongwang1}
\be
S=  \frac{1}{24\pi^2}\int_{w\in W,\mathbb{R}^4} \mathcal{A} \wedge d\mathcal{A}\wedge d\mathcal{A},
\label{eq:14}
\ee
where $w\sim k_x\in W\equiv [-K_x, K_x]$ separates the bulk two Weyl points.
Together with the chiral anomaly from the ``surface" Weyl fermions, and using the gauge field $\mathcal{A} = eA +  w\mathfrak{e}^x$, Eq.~\eqref{eq:14} correctly produces the action ~\eqref{eq:10} in the $\mathcal{V}_a=0$ limit as expected.

%Indeed, the two descriptions are related by a dimensional reduction --- from Eq.~\eqref{eq:9}, by singling out the Berry connection $A_m$ from $\mathcal{A}$ and integrating over $v,u$ directions, one obtains Eq.~\eqref{eq:14}.  %This is similar to the the Teo-Kane classification of defect modes, which only depends on the spatial dimension minus the co-dimension of the defect, but in momentum space.

As an advantage of the $D=4$ description, one can easily take open boundary conditions in the $y$- and/or $z$-directions, since only the momentum direction $w$ is needed in phase space. At a boundary having $z=z_0$, the 4d PSTI has surface states described by a 3d Weyl fermion: $H_{{\rm surf}}=-i \int_{x,y,w} \psi^\dagger \[(\partial_x-iw) \sigma_1 + \partial_y \sigma_2 + \partial_w \sigma_3\] \psi,$ where $w\in [-K_x, K_x],$ and the Weyl fermion is coupled to the background translational gauge field $\mathcal{A}_x = w$ (required to maintain the non-commutativity of $x$ and $k_x$). As we mentioned, the  canonical commutation relation in the $x$-direction is enforced when the spectrum is projected to the low-energy sector, which for Weyl fermions is the cLL. States in the cLL obey the eigenvalue constraint $\sigma_2 \psi = \psi,$ and the effective guiding-center constraint $\langle \partial_x -iw\rangle = 0.$ Thus, projecting $H_{\rm surf}$ to the LLL we find
\be
 \!\!\!\!\!\! H = -i \int_{x,y} \tilde\psi^\dag \partial_y \tilde\psi\textrm{, where }\langle i\partial_x\rangle = \langle w \rangle \in [-K_x,K_x].
 \ee This Hamiltonian represents chiral dispersing surface modes having  $x$-momenta restricted to $[-K_x, K_x]$, i.e., they describe the Fermi arc  of a Weyl semimetal.

As we mentioned before, the PSTI description of GFS's is not unique. Indeed, both descriptions of the WSM (i) $D=6, C=3$ and (ii) $D=4, C=1$ satisfy Eq.~\eqref{eq:1} where $d=3$ and $c=3$. Since the classification of the phase space TI depends only on $c$, the two descriptions are equivalent, and are $\mathbb{Z}$-classified  corresponding to the level of the Chern-Simons actions (\ref{eq:9},\ref{eq:14}). It is well-known that the Chern number is stable even in the presence of interaction effects, and thus the $\mathbb{Z}$ classification applies to interacting systems, as long as translational symmetry is preserved, which is necessary for the definition of phase space and Weyl node separation. For example, it is known that no superconducting order parameter can gap out  a WSM$_2$ (doped or not).~\cite{lihaldane} Furthermore,  even when the Weyl points are gapped via topological order, the $\mathbb{Z}$ classification remains well-defined through the response theory.~\cite{burkov2023}

%WSM with spin-monentum locking. ($\mathbb{Z}_2$ classification in the absence of $U(1)_{\rm spin}$).

\noindent\textbf{$S_4$-symmetric WSM.---}
Recent works have pointed out that more complicated Weyl-point configurations can have quasi-topological responses to electromagnetic and crystalline gauge fields~\cite{dubinkin2021,chongwang1,hirsbrunnerDubBurHug2023}. As an example, we consider a WSM$_4$ having four Weyl nodes arranged in a quadrupolar pattern. Compared with previous works, we further impose a rotoinversion $S_4$ symemtry, a combination between a $C_{4z}$ rotation and a mirror reflection $M_z$, which relates the four Weyl points. Such a semimetal can also be thought of as PSTI, and  interestingly, the corresponding PSTI is a higher-order TI (HOTI) where gapless states are localized at hinges where two surfaces intersect. We will show that the higher-order topology from the bulk has unique signatures for both the surface states and the response theory.

The phase-space coordinates of the HOTI are $(w,v,x,y,z)=(k_x,k_y, x, y, z)$, having surfaces at $w=\pm k_x^0, v=\pm k_y^0,$ and hinges are located at $(w,v)=(\pm k_x^0, \pm k_y^0)$, as illustrated in Fig. \ref{fig:2}(b). From Eq.~\eqref{eq:1}, this HOTI can be constructed starting from a 5d PSTI that has massless 4d Dirac fermions on its surface and then subsequently gapping out the surface states with a spatially dependent mass term $m(w,v)= \mu (v^2-w^2)$ that is odd under $C_4$ rotation in the $(w,v)$ plane while preserving $S_4$~\cite{Schindlereaat_2018}. The hinge states are hence domain wall states of 4d Dirac fermions, i.e., they are are 3d Weyl fermions (bulk Weyl points) (see Fig. \ref{fig:2}(b)).

From Ref.~\cite{qi2008} the effective action of a 5d PSTI is given by
$S=\frac{\Theta}{48\pi^3} \int  \mathcal{F}\wedge \mathcal{F}\wedge \mathcal{F}$, where $\Theta=\pi$, 
and, in analogy with Ref.~\cite{Schindlereaat_2018} a 5d HOTI with $S_4$ symmetry can be described by taking $\Theta$ to be varying (in phase space) as  $\Theta= \pi \sgn (v^2-w^2) + \epsilon$, where $\epsilon\to 0$.
This action is equivalent to Chern-Simons theories on the four side surfaces at $w=\pm k_x^0$ and $v=\pm k_y^0$, each of which has anomalous half-integer levels, i.e.,
\be
S=\sum_{i=1}^{4}\frac{1}{2}\frac{1}{24\pi^2}\int_{{\mathcal{S}_i}\times\mathbb{R}^4} \sgn(v^2-w^2) \mathcal{A}\wedge d\mathcal{A} \wedge d\mathcal{A},
\ee
where $\mathcal{S}_i$'s are the four boundaries in $(w,v)$ forming a square connecting the four Weyl nodes.

As before the background gauge field configuration to leading order in momenta is given by $\mathcal{A} = eA + w \mathfrak{e}^x+v \mathfrak{e}^y$. Since the $k$-space dipole moment of the Weyl point vanishes, the lowest order response to the momentum gauge field is quadratic in momenta. We obtain 
\begin{align}
S=&\sum_{i=1}^{4}\frac{e}{2}\frac{(-1)^i}{8\pi^2}\int_{{\mathcal{S}_i}\times\mathbb{R}^4}  d\(k_\alpha \mathfrak{e}^\alpha\) \wedge d\(k_\beta \mathfrak{e}^\beta\) \wedge A \nonumber\\
%= & \frac{1}{2}\frac{1}{8\pi^2}\int_{\mathbb{R}^4} A'^{(0)}\wedge dA'^{(0)} \wedge A \Big|_{(-k_x^0,k_y^0)}^{(k_x^0,k_y^0)}  \nonumber\\
= & \frac{e\mathcal{Q}_{\alpha\beta}}{8\pi^2}\int_{\mathbb{R}^4} \mathfrak{e}^\alpha \wedge d\mathfrak{e}^\beta \wedge A ,
\end{align}
 where $\mathcal{Q}_{\alpha\beta}$ is the $k$-space quadrupole moment formed by the Weyl points.  In our case $\mathcal{Q}_{xy} =\mathcal{Q}_{yx} = 4k_x^0 k_y^0$. By a boundary anomaly argument similar to the WSM$_2$ case above, one can show that $\mathcal{Q}_{\alpha\beta}$ remains unchanged upon doping as long as the dispersion around the Weyl points is linear. This is precisely the response theory in Refs.~\cite{chongwang1,hirsbrunnerDubBurHug2023}  (albeit derived using other methods, and for a WSM$_4$ where $\mathcal{Q}_{\alpha\beta}$ is rotated by 45-degrees), which results in fractional charges bound at dislocations. 
 Note that in the absence of $S_4$ symmetry, after integrating over momentum space there is in general  another term in the action $\propto \mathfrak{e}^x \wedge \mathfrak{e}^y \wedge dA$, which corresponds to a non-vanishing charge polarization in the $z$-direction. While this term was omitted in previous works without justification, for the present system its coefficient vanishes (modulo quantized contributions from filled bands) since it is odd under  $S_4$ symmetry. 

The surface states for this system can also be directly obtained from phase-space topology. The top surface at $z=z_0$ is parameterized by $(w,v, x,y)$ and harbors a 4d surface Dirac fermion subject to mass domain walls,
\begin{align}
H_{\rm surf} =& -i \int_{w,v,x,y} \psi^\dag \[(\partial_x-iw) \Gamma_1 + (\partial_y -iv) \Gamma_2  \right. \nonumber\\
&\left. + \partial_w\Gamma_3+\partial_v \Gamma_4+ i \mu (v^2-w^2) \Gamma_5\] \psi,
\end{align}
where $\{\Gamma_{i}\}$ are anticommuting $\Gamma$ matrices with $\Gamma_5=\Gamma_1\Gamma_2\Gamma_3\Gamma_4$, and we have added the background gauge field  $\mathcal{A}_x = w$ and $\mathcal{A}_y=v$ to preserve the non-commutativity of $x,k_x$ and $y,k_y$ respectively. The two orthogonal phase-space magnetic fields allow us to project the Dirac fermion to the zeroth Landau level via $i\Gamma_1\Gamma_3 \psi = i\Gamma_2\Gamma_4 \psi = \psi,$ and thus $\Gamma_5\psi = \psi$. Such a projection quenches the dispersion in the $w,v$ directions and enforces the guiding center relations $\langle \partial_x -iw\rangle =\langle \partial_y -iv\rangle = 0.$ The remaining mass term proportional to $\Gamma^5$ hence projects to: 
\be
H = \int_{x,y} \psi^\dag (\partial_x^2-\partial_y^2)\psi,
\ee
which is a saddle-point that forms crossing Fermi arcs connecting the four Weyl points.  Interestingly, such a dispersion was recently shown to exhibit rank-2 chiral anomaly~\cite{dubinkin2021,zhu2023higher}, which can be canceled from the bulk using the response action we derived above.

\noindent\textbf{Outlook.---} In this work we presented a unified framework of treating GFS's as phase-space TIs, leading naturally to non-perturbative results such as generalized Luttinger relations, topological response theories, and surface state properties. As we mentioned, one immediate future direction  is the classification of interacting GFS's from that of  PSTIs. It will also be interesting to extend this framework to GFSs with topological orders.

\begin{acknowledgments}
\noindent\textbf{Acknowledgments.---} We would like to thank Yi-Hsien Du, Lei Yang, and Chong Wang for useful discussions. YW is supported by NSF under award number DMR-2045781. TLH thanks ARO MURI W911NF2020166 for support. We acknowledge support by grant NSF PHY-1748958 to the Kavli Institute for Theoretical Physics (YW and TLH), and by grant NSF PHY-2210452 to the Aspen Center for Physics (YW), where this work was initiated and performed.

\noindent{\textbf{Notes added.--- }After submitting this manuscript to arXiv, we became aware of a related work by Wang and Sau~\cite{wangsau2024} that was announced the same day.}

\end{acknowledgments}

\bibliography{biblio}

\clearpage
\newpage
%%%%%%%% commands for reset equation, figure and refs numbering in supplementary material
\renewcommand{\theequation}{S\arabic{equation}}
\renewcommand{\thefigure}{S\arabic{figure}}
\renewcommand{\bibnumfmt}[1]{[S#1]}
%\renewcommand{\citenumfont}[1]{S#1}
%%%%%%%%%%%%%%%%%%%%%%%%%%%%%%
\setcounter{equation}{0}
\setcounter{figure}{0}
\setcounter{page}{1}

\onecolumngrid
%~\\[40pt]
\title {Supplemental Material Gapless Fermionic Systems as Phase-space Topological Insulators: Non-perturbative Results from Anomaly}

\date{\today}

\begin{abstract}
In this Supplemental Material we derive in details the response theories of a doped 2d Dirac semimetal with two Dirac points (DSM$_2$) and a doped Weyl semimetal with two Weyl points (WSM$_2$) from the phase-space TI correspondence.
\end{abstract}
\maketitle

\onecolumngrid

\section{doped DSM$_2$}
As we mentioned in Table I of the main text, a doped Dirac semimetal with two Dirac points (DSM$_2$) in 2d can be described via its low-energy properties around the FS if Berry phase effects as included. As we explained in the main text, 2d FSs correspond to the surface states of a 4d Chern insulator in phase space. In particular, we consider the configuration in which a 4d Chern insulator surrounds two circular electron pockets $\mathcal{U}_{1,2}$ of trivial insulators in the $(w,v)\sim (k_x,k_y)$ directions having radius $K_F$. To describe the Berry phase effects, there exists a $\pi$ flux in each of the trivial insulator regions, which corresponds to two Dirac points enclosed in the two FS's. Without loss of generality, we assume the two Dirac points are at $\vec k^{(1,2)}$.

The response theory for the 4d phase-space Chern insulator is given by~\cite{qi2008}
\be
S = \frac{1}{24\pi^2}\int_{(w,v)\in \mathrm{BZ}, x,y, t} \mathcal{A}\wedge d\mathcal{A} \wedge d\mathcal{A} -  \sum_{a=1}^2\frac{1}{24\pi^2}\int_{(w,v)\in \mathcal U_{a}, x,y, t} \mathcal{A}\wedge d\mathcal{A} \wedge d\mathcal{A},
\label{eq:0}
\ee
where BZ denotes the entire Brillouin zone. The first term overcounts the phase-space volume of the 4d PSTI and the second term corrects it. The phase-space background gauge field can be decomposed as
\be
\mathcal{A} = eA + k_i\mathfrak{e}^i + \mathfrak{a},
\ee
where $A$ is the electromagnetic field in real space, $k_i\mathfrak{e}^i$ is the translation gauge field where $k=(w,v)$, and $\mathfrak{a}$ is the Berry connection. Over a contour that encloses a Dirac point, we have $\oint \mathfrak{a} = \pm \pi$, and 
\be
\int_{-\pi}^{\pi} dv \mathfrak{a}_v = \pi \theta(w^2-k_0^2).
\label{eq:3}
\ee

From the first term of Eq.~\eqref{eq:0}, at linear order in $A$, we get two terms. The first is obtained by taking $k_i\mathfrak{e}^i$ components for all $\mathcal{A}$'s but one:
\be
S_1= - \frac{e\mathcal{V}_{\rm BZ}}{4\pi^2} \int_{x,y,t} A\wedge \mathfrak{e}^x \wedge \mathfrak{e}^y.
\ee
The second comes from mixing $k_i\mathfrak{e}^i$ and $\mathfrak{a}$ by noting 
\be
\int_{(w,v)\in \mathrm{BZ}} dk_i\wedge \mathfrak{a} = -\pi \mathcal{P}_{0,i},
\ee
where $\mathcal{P}_{0,i}$ is the $i$-th component of the dipole moment formed by the two Dirac points in momentum space, i.e.,
\be
{\mathcal{\vec P}_0} = \vec k^{(1)} - \vec k^{(2)}.
\ee
We have
\begin{align}
S_2 = &\frac{1}{4\pi^2}\int_{w,v,x,y,t} \mathfrak{a}\wedge (dk_i \mathfrak{e}^i) \wedge (edA) \nonumber\\
=&\frac{e\mathcal{P}_{0,i}}{4\pi} \int_{x,y,t} \mathfrak{e}^i\wedge dA.
\end{align}

At quadratic order in $A$, the contribution to the first term of Eq.~\eqref{eq:0} comes from taking $\mathfrak{a}$ components for one of the $\mathcal{A}$'s. Using Eq.~\eqref{eq:3} we obtain
\be
S_3= \frac{e^2}{8\pi} \int_{w\in W, x, y,t} dA\wedge dA,
\ee
where $W\equiv [-k_0, k_0]$. Since the background values of $dA$ have only spacetime components, this integral vanishes.

From the second term of Eq.~\eqref{eq:0}, there are two terms that are linear in $A$. Analogous to $S_1$, we have
\be
S_4=\sum_{a=1}^2 \frac{e\mathcal{V}_{a}}{4\pi^2} \int_{x,y,t} A\wedge \mathfrak{e}^x \wedge \mathfrak{e}^y.
\ee
Analogous to $S_2$, we have
\be
S_5 = - \sum_{a=1}^2 \frac{1}{4\pi^2}\int_{(w,v)\in \mathcal{U}_a,x,y,t} \mathfrak{a}\wedge d(k_i \mathfrak{e}^i) \wedge (edA),
\ee
which we can rewrite as 
\begin{align}
S_5 = &\frac{e\Delta\mathcal{P}_{i}}{4\pi} \int_{x,y,t} \mathfrak{e}^i\wedge dA,
\end{align}
where 
\be
\Delta \mathcal{P}_{i} = - \sum_{a=1}^2 \int_{(w,v)\in \mathcal U_a} \frac{\mathfrak{a}}{\pi}\wedge dk_i = \sum_{a=1}^2 \[\int_{\mathrm{FS}_a} \frac{\mathfrak{a}}{\pi} k^{(a)}_{F,i} -\int_{(w,v)\in \mathcal U_a} \frac{d \mathfrak{a}}{\pi} k^{(a)}_i\] = \sum_{a=1}^2 \int_{\mathrm{FS}_a} \frac{\mathfrak{a}}{\pi} k^{(a)}_{F,i} - \(k^{(1)}_{i} - k^{(2)}_i\),
\ee
where in the second step we integrated by parts and used Stokes' theorem, and in the last step we used the fact that for a Dirac point $d\mathfrak{a}(\vec k)=\pm \pi \delta(\vec k-\vec k^{(a)}) dw\wedge dv$, where $\vec k^{(a)}$ is the location of the $a$-th Dirac point. Combining $S_2$ and $S_5$, it is convenient to define
\be
\mathcal{P}_i \equiv  \mathcal{P}_{0,i} + \Delta \mathcal{P}_i =  \sum_{a=1}^2 \int_{\mathrm{FS}_a} \frac{\mathfrak{a}}{\pi} k^{(a)}_{F,i}.
\label{eq:121}
\ee

As an approximation, we assume the fermionic dispersion inside $\mathcal{V}_a$ is linear as a function of momentum (which is assured for small enough doping). Consequently, the Berry connection inside $\mathcal{V}_a$ is an odd function of the momentum  deviation from the Dirac points. In this case, we have
\be
\int_{(w,v)\in \mathcal{V}_a}  \mathfrak{a} \wedge dk_i=0,~~~ \Delta \mathcal{P}_i=0.
\ee
Under this approximation $\mathcal{\vec P} =\mathcal{\vec P}_{0} = \vec k^{(1)} - \vec k^{(2)}$. In other words, the presence of Fermi pockets does not affect the coefficient $\mathcal{P}$ for small enough doping that the dispersion is roughly linear.

At quadratic order in $A$, there is a contribution from the second term that is similar to $S_3$, given by
\begin{align}
S_6=& -\sum_{a=1}^2\frac{e^2}{8\pi^2} \int_{k_r^{(a)}\in K_F , x, y,t} dA\wedge dA\int_{S_r^{(a)}} \mathfrak{a} \nonumber \\
 = &\sum_{a=1}^2\frac{(-)^ae^2}{8\pi} \int_{k_r^{(a)}\in K_F , x, y,t} dA\wedge dA.
\end{align}
Again, since the background values of $dA$ have only spacetime components, this integral vanishes.

Combining terms from $S_1$ to $S_6$, we obtain
\begin{align}
S=&\int_{x,y,t} \left[ \frac{e\mathcal{ V}}{4\pi^2} A\wedge \mathfrak{e}^x \wedge \mathfrak{e}^y +\frac{e\mathcal {P}_i}{4\pi} \mathfrak{e}^i\wedge dA \right].
\label{eq:dirac}
\end{align}
where $\mathcal{V}$ is the combined volume enclosed by the Fermi surfaces, and, as we mentioned, $\mathcal{\vec P}$ is the dipole moment formed by the Dirac points.

In Eq.~\eqref{eq:dirac}, the first term describes Luttinger's theorem. Indeed, making use of the background values $\mathfrak{e}^i_j=
\delta^i_j$, we see that
\be
N = \int \frac{\delta S}{e\delta A_0} = \frac{\mathcal{V}}{4\pi^2}\int dx \wedge dy = \frac{\mathcal{V}V}{4\pi^2},
\ee
where $V=\int dx dy$ is the (position-space) area of the sample. The second term leads to the electric polarization
\be
p_i = \int \frac{\delta S}{e\delta (\partial_0A_i - \partial_i A_0)} = \frac{e\epsilon_{ik}  \mathcal{P}_j\mathfrak{e}^j_k}{4\pi}    \int dx \wedge dy=  \frac{\epsilon_{ij}  \mathcal{P}_jV}{4\pi}.
\ee

We show that the value of $\mathcal{P}_{i}$ can be nonperturbatively established directly from the perspective of boundary anomaly, which is equivalent with the bulk approach taken above. To this end, we begin by noting that from the bulk theory \eqref{eq:dirac} there is a  momentum current response to the translation gauge field $\mathfrak{e}^i$ via $\mathfrak{j}_i^\mu\equiv \delta S/\delta \mathfrak{e}^i_\mu$, which carries the $i$-th component of momentum  in the $\mu$-th direction. The total momentum is $\langle P_i\rangle=\int_{\mathbb{R}^2}\star \mathfrak{j}_i.$ We get:
\begin{align}
\partial_t\langle P_i\rangle dt =& \int_{\mathbb{R}^2} d\star \mathfrak{j}_i= \int_{\mathbb{R}^2}  \[\frac{e\mathcal V}{4\pi^2}\epsilon_{ij} F\wedge \mathfrak{e}^j + \frac{e^2 }{4\pi}d\mathcal{P}_{i}\wedge F \],
\label{eq:160}
\end{align} where $F=dA.$
Together with  Luttinger's theorem, we see the first term on the right hand side captures the electric force on the system. The second term corresponds to the Lorentz force on the system when the polarization changes, e.g., by turning on a mass term to gap out the Dirac point, causing an electric current that couples to a magnetic field.

As we mentioned, the same result can also be obtained from an emergent anomaly of the PSTI boundary. From the PSTI analogy, the gapless fermions on the $a$-th FS exhibit the chiral anomaly for the charge current in a (3+1)d phase space $(x,y,\theta, t)$, where $\theta$ parameterizes the FS,
\be
d\star j^{(a)}(\vec k_F^{(a)}) = \frac{e}{4\pi^2}F\wedge d(k_j \mathfrak{e}^j)+\frac{e}{4\pi^2} F\wedge \mathfrak{f}_m(\vec k_F^{(a)}).
\label{eq:170}
\ee
The FS contribution to the momentum current depends on the particle current and is given by $\mathfrak{j}^{(a)}_i \equiv \oint_{{\rm FS}_a}  k_{F,i}^{(a)}\, j^{(a)}(\vec k_F^{(a)})$. Using the fact that $\mathfrak{a}_m$ depends only on momentum and is independent of spatial coordinates, we integrate over $\mathbb{R}^2\times \mathrm{FS}_a$ and find:
\be
\partial_t\langle P_i\rangle dt = \sum_{a=1}^2\int_{\mathbb{R}^2} d\star \mathfrak{j}_i^{(a)}= \int_{\mathbb{R}^2}  \frac{e\mathcal V}{4\pi^2}\epsilon_{ij} F\wedge \mathfrak{e}^j + \sum_{a=1}^2 \oint_{\mathrm{FS}_a}k_{F,i}^{(a)} \frac{d\mathfrak{a}_m}{\pi}\int_{\mathbb{R}^2}  \frac{e^2}{4\pi} F.
\label{eq:19}
\ee
 To obtain the  the first term of the right hand side we used Stokes' theorem to convert the FS integral into the volume $\mathcal{V}$ enclosed by the FS. Matching \eqref{eq:160} with \eqref{eq:19} immediately leads to
\be
\mathcal{P}_{i} = \sum_{a=1}^2 \oint_{\mathrm{FS}_a}k_{F,i}^{(a)} \frac{\mathfrak{a}}{\pi},
\ee
which agrees with Eq.~\eqref{eq:121} obtained from a bulk approach.
This result is the analog of Eq.~(16) of the main text, i.e., electric polarization  for a doped Dirac semimetal is completely a FS phenomenon. For linear dispersion around Dirac points, as we mentioned, $\mathcal{P}_{i}$ reduces to the dipole moment formed by the two Dirac points in momentum space, i.e., $\mathcal{P}_{0,i}= k_{1,i} - k_{2,i}.$

\section{doped WSM$_2$}
We begin by considering a phase space configuration in which two trivial insulators fill regions $\mathcal{V}_1$ and $\mathcal{V}_2$ in $(w,v,u)\sim(k_x,k_y,k_z)$ coordinates corresponding to the hole-like Fermi pockets around each Weyl point at $\vec k^{(1,2)}$ with chirality $\chi_{1,2}=\mp 1$. In addition, there is a magnetic monopole in $(w,v,u)$ inside each of the Fermi pockets, corresponding to the Weyl nodes. 

To illustrate an interesting parallel between doped DSM$_2$ and doped WSM$_2$, we have used same notations such as $\mathcal{P}_i$, $k^{(a)}$, $\mathcal{V}$ for both systems. We caution that they should however not be confused, e.g., $\mathcal{V}$ is the volume inside a 3D Fermi surface, not the area inside a 2D Fermi surface as in the previous section.

From phase-space geometry the action can be equivalently written as~\cite{qi2008}
\begin{align}
S=&\frac{1}{192\pi^3}\int_{(w,v,u)\in \mathrm{BZ},x,y,z,t} \mathcal{A}\wedge d\mathcal{A}\wedge d\mathcal{A}\wedge d\mathcal{A} \nonumber\\
&-\sum_a\frac{1}{192\pi^3}\int_{(w,v,u)\in \mathcal{V}_a,x,y,z,t} \mathcal{A}\wedge d\mathcal{A}\wedge d\mathcal{A}\wedge d\mathcal{A},
\label{eq:1}
\end{align}
where the first term overcounts the volume of TI and the second term corrects for it. We decompose the gauge field as 
\be
\mathcal{A} = eA + k_i\mathfrak{e}^i + \mathfrak{a}_m,
\ee
where $A$ is the electromagnetic field in real space, $k_i\mathfrak{e}^i$ is the translation gauge field with $k_i=(w,v,u)_i$ and $\mathfrak{e}^i_j=\delta^i_j$, and $\mathfrak{a}_m$ is Berry connection created by the two monopoles in $\vec k$-space, which over a  surface enclosing one of the monopoles satisfies $\oint \nabla\times \mathfrak{a}_m = \pm 2\pi$.

Electromagnetic response can be obtained by integrating over background configurations for $k_i\mathfrak{e}^i$ and $\mathfrak{a}_m$ along momentum directions. In the first term of Eq.~\eqref{eq:1}, a response to linear order in $A$ can be obtained by taking three $\mathcal A$ components to be those of the translational gauge fields, and by using 
\be
d(k_i\mathfrak{e}^i)= dw\wedge\mathfrak{e}^x + dv\wedge\mathfrak{e}^y + du\wedge\mathfrak{e}^z.
\ee
We have
\begin{align}
S_1=& \frac{1}{8\pi^3}\int_{(w,v,u)\in \mathrm{BZ},x,y,z,t} eA \wedge \(dw\wedge\mathfrak{e}^x\) \wedge \(dv\wedge\mathfrak{e}^y\) \wedge \(du\wedge \mathfrak{e}^z\) \nonumber\\
=& \frac{e\mathcal{V}_{\rm BZ}}{8\pi^3}\int_{x,y,z,t} A \wedge \mathfrak{e}^x\wedge \mathfrak{e}^y\wedge \mathfrak{e}^z.
\end{align}
At quadratic order in $A$, the contribution to the electromagnetic response can be obtained by noticing
\be
\int_{(w,v,u)\in \mathrm{BZ}} dk_i\wedge (d\mathfrak{a}_m) = - 2\pi \mathcal{P}_{0,i},
\ee
where 
\be
\mathcal{P}_{0,i} = \sum_{a=1}^2 k_i^{(a)} \chi_a
\ee
($\chi_a$ being the chirality of the Weyl point) is the  $i$-th component of the polarization of the Weyl points. This is because the Weyl points in momentum space separate layers with Chern numbers that differ by 1. This gives rise to 
\begin{align}
S_2=& \frac{1}{16\pi^3}\int_{w,v,u,x,y,z,t} eA \wedge \(e dA\) \wedge \(dk_i\wedge\mathfrak{e}^i\) \wedge \(d\mathfrak{a}_m\). \nonumber\\
=& -\frac{e^2\mathcal {P}_{0,i}}{8\pi^2} \int_{x,y,z,t} \mathfrak{e}^i\wedge A\wedge dA.
\end{align}
{The contribution at cubic order in $A$ vanishes for background values of $A$, which we do not include here.}

From the second term in Eq.~\eqref{eq:1}, there is first a contribution at linear order in $A$. Similar to what we did for the first term, we have
\begin{align}
S_4=& -\sum_a\frac{1}{8\pi^3}\int_{(w,v,u)\in \mathcal{V}_a,x,y,z,t} eA \wedge \(dw\wedge\mathfrak{e}^x\) \wedge \(dv\wedge\mathfrak{e}^y\) \wedge \(du\wedge \mathfrak{e}^z\) \nonumber\\
=& -\sum_a\frac{e\mathcal{V}_{a}}{8\pi^3}\int_{x,y,z,t} A \wedge \mathfrak{e}^x\wedge \mathfrak{e}^y\wedge \mathfrak{e}^z.
\end{align}
At quadratic order in $A$, the contribution from the second term in Eq.~\eqref{eq:1} is given by
\begin{align}
S_5=& -\sum_a\frac{1}{16\pi^3}\int_{(w,v,u)\in \mathcal{V}_a.x,y,z,t} eA \wedge \(e dA\) \wedge \(dk_i\wedge\mathfrak{e}^i\) \wedge \(d\mathfrak{a}_m\),
\end{align}
which we can rewrite as 
\begin{align}
S_5 = &\frac{e^2\Delta\mathcal {P}_{i}}{8\pi^2} \int_{x,y,z,t} \mathfrak{e}^i\wedge A\wedge dA.
\end{align}
where by similar steps as in  Eq.~\eqref{eq:121}, we have
\be
\Delta \mathcal{P}_{i} = - \sum_{a=1}^2 \int_{(w,v)\in \mathcal U_a} \frac{d\mathfrak{a}_m}{2\pi}\wedge dk_i = -
\sum_{a=1}^2\[\oint_{{\rm FS}_a}  
k_{F,i}^{(a)}\frac{\mathfrak{f}_m(\vec k_F^{(a)})}{2\pi} -\chi_a k_i^{(a)}\],
\ee
where $\mathfrak{f}_m\equiv d\mathfrak{a}_m$. Comparing $S_2$ and $S_5$, it is convenient to define
\be
\mathcal{P}_i \equiv  \mathcal{P}_{0,i} + \Delta \mathcal{P}_i = -
\sum_a\oint_{{\rm FS}_a}  
k_{F,i}^{(a)}\frac{\mathfrak{f}_m(\vec k_F^{(a)})}{2\pi}.
\label{eq:122}
\ee
This derivation is complementary to the same result we found in Eq.~(16) of the main text, which was derived from the boundary anomaly.

For small $\mathcal{V}_a$ within which the dispersion around the Weyl points is approximately linear, the Berry curvature is an odd function of the momentum deviation from the Weyl point, and thus
\be
\int_{(w,v,u)\in \mathcal{V}_a}  (d\mathfrak{a}_m) \wedge dk_i = 0,
~~~\Delta\mathcal{P}_i=0.
\ee
In other words, to leading order the size of Fermi pockets does not contribute to a correction to the coefficient $\mathcal{P}_{i}$, and $\mathcal{P}_{i}=\mathcal{P}_{0,i}$.

Combining all terms, we obtain Eq.~(9) of the main text:
\begin{align}
S=&\int_{x,y,z,t} \left[ \frac{e\mathcal{V}}{8\pi^3} A\wedge \mathfrak{e}^x \wedge \mathfrak{e}^y \wedge \mathfrak{e}^z -\frac{e^2\mathcal {P}_i}{8\pi^2} \mathfrak{e}^i\wedge A\wedge dA \right] 
\label{eq:10sm}
\end{align}
where $\mathcal{ V}= \mathcal{V}_{\rm BZ}- \sum_{a=1,2}\mathcal{V}_{1,2}$ is the volume enclosed by the FS's. As we mentioned in the main text, the first term in \eqref{eq:10sm} leads to the Luttinger theorem, while the second term is responsible for the anomalous Hall effect.

%\section{Analog of the Witten effect in a 6d phase-space TI}

%\bibliography{biblio}

%\end{widetext}

\end{document}